\documentclass
[preprint,showpacs,superscriptaddress,prc]{revtex4}%
\usepackage{amsfonts}
\usepackage{amsmath}
\usepackage{amssymb}
\usepackage{graphicx}%
\begin{document}
\preprint{JLAB-THY-05-400}
\preprint{NT@UW-05-11}
\title{Study of Lattice QCD \ Form Factors Using the Extended Gari-Kr\"{u}mpelmann Model}
\author{Hrayr H. Matevosyan}
\affiliation{Louisiana State University, Department of Physics \&
              Astronomy, 202 Nicholson Hall, Tower Dr., LA 70803, USA}
 \affiliation{Thomas Jefferson National Accelerator Facility, 12000
              Jefferson Ave., Newport News, VA 23606, USA}
 \affiliation{Laboratory of Theoretical Physics, JINR, Dubna, Russian
              Federation}
\author{Anthony W. Thomas}
 \affiliation{Thomas Jefferson National Accelerator Facility, 12000
              Jefferson Ave., Newport News, VA 23606, USA}
\author{Gerald A. Miller}
\affiliation{University of Washington, Department of Physics, Box 351560, Seattle, WA
98195-1560, USA}

\begin{abstract}
We explore the suitability\ of a modern vector meson dominance (VMD) model as
a method for chiral extrapolation of nucleon electromagnetic form factor
simulations in lattice QCD. It is found that the VMD fits to experimental data
can be readily generalized to describe the lattice simulations. However, the
converse is not true. That is, the VMD form is unsuitable as a method of
extrapolation of lattice simulations at large quark mass to the physical regime.

\end{abstract}
\date{08/24/2005}
\startpage{1}
\endpage{14}
\maketitle

\section{Introduction}

The electromagnetic form factors of the nucleon are a major source of
information about its internal structure. On the experimental side, the unique
capabilities of Jefferson Lab have recently led to a major revision of our
knowledge of the proton electric form factor, with $G_{E}/G_{M}$ unexpectedly
decreasing quite rapidly with increasing $Q^{2}~$%
\cite{Jones:1999rz,Gayou:2001qd}. Although various models of hadron structure
have reasonable success in the low-$Q^{2}$ regime, only a few have any claim
to also describing the\ $Q^{2}$ dependence of $G_{E}/G_{M}$ while reproducing
the magnitude of $G_{M}$ - see the review \cite{Hyde-Wright:2004gh}. There is
no consensus as to which model best represents QCD. However, at least one of
them, the Light Front Cloudy Bag Model (LFCBM)
\cite{Miller:2002ig,Miller:2002qb}, did indeed anticipate the behavior found
at JLab.

Lattice QCD\ has the great attraction of being able to give us the unambiguous
consequences of non-perturbative QCD. Until now it has proven possible to
calculate proton and neutron form factors to $Q^{2}$ of order 3 $%
\operatorname{GeV}%
^{2}$ in quenched QCD \cite{Gockeler:2003ay}. While this data is restricted to
relatively large pion mass ($m_{\pi}>$ $0.5%
\operatorname{GeV}%
$), it is remarkable that a model like the LFCBM is able to not only describe
the experimental data, but with relatively mild assumptions about the mass
dependence of 2 parameters, it also produces an excellent description of the
lattice QCD form factor data \cite{Matevosyan:2005fz}.

The latter finding works two ways. First it assures us that the model passes a
test that any acceptable model should pass~\cite{Detmold:2001hq}, namely that
it is consistent with the dependence of the quark mass found in QCD itself.
Second, as there is no model independent way for chiral extrapolation of
hadron properties at high-$Q^{2}$, it suggests that the LFCBM presents a
reasonable method whereby one can extrapolate hadron properties found at large
light quark mass to the physical pion mass.

Given the considerable interest in the vector meson dominance (VMD) approach,
in this paper we consider its suitability as a method of chiral extrapolation.
There are many variations of the basic VMD, but we choose the implementation
of Lomon \cite{Lomon:2002jx}, because it is phenomenologically extremely
successful. We introduce the dependence of the vector meson masses in $m_{\pi
}$ found in earlier lattice studies \cite{Leinweber:1998ej,RHO_MASS} and
parametrize the mass dependence of the corresponding couplings in order to
best describe the lattice QCD data. The major disappointment is that the
functional form is so sensitive to the parameters, that it is meaningless to
compare with extrapolated data any extrapolation of the form factors to the
physical pion mass. This makes the VMD\ approach unsuitable as a method of
chiral extrapolation. On the other hand, by fixing the parameters to the
values given in Ref. \cite{Lomon:2002jx} at the physical mass, it is possible
to obtain a fit to all of the lattice QCD data of comparable quality to that
found earlier using LFCBM.

\section{Lattice Data Fits and Results}

We employ the Extended Gari-Kr\"{u}mpelmann Model (GKex) of Lomon
Ref.\cite{Lomon:2002jx}\ to fit Lattice QCD calculated nucleon electric and
magnetic form factors produced by the QCDSF collaboration
\cite{Gockeler:2003ay}.

\subsection{Review of the GKex Model}

Here we briefly summarize the formulation of the GKex model from
Ref.\cite{Lomon:2002jx}. The extended Gari-Kr\"{u}mpelmann model exhibits the
basic properties of a VMD\ model, and also phenomenologically incorporates the
correct high-$Q^{2}$ behavior of the nucleon electromagnetic form factors as
implied by PQCD. The model was successfully fit to the present experimental
data sets available for the nucleon electromagnetic form factors. The
particular interest in the model is increased by its ability to describe the
fall-off of the proton ratio, $G_{E}/G_{M}$, vs $Q^{2}$, as measured recently
in Refs.\cite{Jones:1999rz,Gayou:2001qd}.

Our goal is to calculate the Dirac, $F_{1}$, and Pauli, $F_{2}$, form factors,
defined through the nucleon electromagnetic current as:
\begin{equation}
\left\langle N,\lambda^{\prime}p^{\prime}\left\vert J^{\mu}\right\vert
N,\lambda p\right\rangle =\overline{u}_{\lambda^{\prime}}(p^{\prime})\left[
F_{1}(Q^{2})\gamma^{\mu}+\frac{F_{2}(Q^{2})}{2M_{N}}i\sigma^{\mu\nu}%
(p^{\prime}-p)_{\nu}\right]  u_{\lambda}(p)\,. \label{FM_EM_CURRENT}%
\end{equation}
The momentum transfer is $q^{\mu}=($ $p^{\prime}-p)^{\mu},~Q^{2}=-q^{2}$ and
$J^{\mu}$ is taken to be the electromagnetic current operator for a nucleon.
For $Q^{2}=0$ the form factors $F_{1}$ and $F_{2}$ are, respectively, equal to
the charge and the anomalous magnetic moment, $\kappa$, in units of $e$ and
$e/(2M_{N})$, while the magnetic moment is $\mu=F_{1}(0)+F_{2}(0)=1+\kappa$.

We are interested in the electric and magnetic Sachs form factors, which are
defined as%
\begin{equation}
G_{E}=F_{1}-\frac{Q^{2}}{4M_{N}^{2}}F_{2},\hspace{2cm}G_{M}=F_{1}+F_{2}
\label{FM_G_TO_F}%
\end{equation}
with normalization
\begin{align}
G_{E}^{p}(0)  &  =1;\label{FM_FF_NORM}\\
G_{M}^{p}(0)  &  =\mu_{p};\nonumber\\
G_{E}^{n}(0)  &  =0;\nonumber\\
G_{M}^{n}(0)  &  =\mu_{n};\nonumber
\end{align}

One can express the Pauli and Dirac form factors in terms of isoscalar and
isovector form factors
\begin{align}
2F_{i}^{p}  &  =F_{i}^{IS}+F_{i}^{IV};\label{FM_DP_TO_ISIV}\\
2F_{i}^{n}  &  =F_{i}^{IS}-F_{i}^{IV};\nonumber
\end{align}

The isoscalar and isovector form factors were parametrized by Lomon as%
\begin{align}
F_{1}^{iv}(Q^{2})  &  =N/2\frac{1.0317+0.0875(1+Q^{2}/0.3176)^{-2}}%
{(1+Q^{2}/0.5496)}F_{1}^{\rho}(Q^{2})\label{FM_ISIV_GKEX}\\
&  \qquad{}+\frac{g_{\rho^{\prime}}}{f_{\rho^{\prime}}}\frac{m_{\rho^{\prime}%
}^{2}}{m_{\rho^{\prime}}^{2}+Q^{2}}F_{1}^{\rho}(Q^{2})+\left(
1-1.1192\,N/2-\frac{g_{\rho^{\prime}}}{f_{\rho^{\prime}}}\right)  F_{1}%
^{D}(Q^{2})\nonumber\\
F_{2}^{iv}(Q^{2})  &  =N/2\frac{5.7824+0.3907(1+Q^{2}/0.1422)^{-1}}%
{(1+Q^{2}/0.5362)}F_{2}^{\rho}(Q^{2})\nonumber\\
&  \qquad{}+\kappa_{\rho^{\prime}}\frac{g_{\rho^{\prime}}}{f_{\rho^{\prime}}%
}\frac{m_{\rho^{\prime}}^{2}}{m_{\rho^{\prime}}^{2}+Q^{2}}F_{2}^{\rho}%
(Q^{2})+\left(  \kappa_{\nu}-6.1731\,N/2-\kappa_{\rho^{\prime}}\frac
{g_{\rho^{\prime}}}{f_{\rho^{\prime}}}\right)  F_{2}^{D}(Q^{2})\nonumber\\
F_{1}^{is}(Q^{2})  &  =\frac{g_{\omega}}{f_{\omega}}\frac{m_{\omega}^{2}%
}{m_{\omega}^{2}+Q^{2}}F_{1}^{\omega}(Q^{2})+\frac{g_{\omega^{\prime}}%
}{f_{\omega^{\prime}}}\frac{m_{{\omega^{\prime}}}^{2}}{m_{\omega^{\prime}}%
^{2}+Q^{2}}F_{1}^{\omega}(Q^{2})+\frac{g_{\phi}}{f_{\phi}}\frac{m_{\phi}^{2}%
}{m_{\phi}^{2}+Q^{2}}F_{1}^{\phi}(Q^{2})\nonumber\\
&  \qquad{}+\left(  1-\frac{g_{\omega}}{f_{\omega}}-\frac{g_{\omega^{\prime}}%
}{f_{\omega^{\prime}}}\right)  F_{1}^{D}(Q^{2})\nonumber\\
F_{2}^{is}(Q^{2})  &  =\kappa_{\omega}\frac{g_{\omega}}{f_{\omega}}%
\frac{m_{\omega}^{2}}{m_{\omega}^{2}+Q^{2}}F_{2}^{\omega}(Q^{2})+\kappa
_{\omega^{\prime}}\frac{g_{\omega^{\prime}}}{f_{\omega^{\prime}}}%
\frac{m_{\omega^{\prime}}^{2}}{m_{\omega^{\prime}}^{2}+Q^{2}}F_{2}^{\omega
}(Q^{2})+\kappa_{\phi}\frac{g_{\phi}}{f_{\phi}}\frac{m_{\phi}^{2}}{m_{\phi
}^{2}+Q^{2}}F_{2}^{\phi}(Q^{2})\nonumber\\
&  \qquad{}+\left(  \kappa_{s}-\kappa_{\omega}\frac{g_{\omega}}{f_{\omega}%
}-\kappa_{\omega^{\prime}}\frac{g_{\omega^{\prime}}}{f_{\omega^{\prime}}%
}-\kappa_{\phi}\frac{g_{\phi}}{f_{\phi}}\right)  F_{2}^{D}(Q^{2})\nonumber
\end{align}
with pole terms of the $\omega(782)$, $\phi(1020)$, $\omega^{\prime}(1420)$ ,
$\rho(770)$ and $\rho^{\prime}(1450)$ mesons, and the $F_{i}^{D}$ terms
ensuring the correct asymptotic behavior as calculated in PQCD. The
$F_{i}^{\alpha}$, with $\alpha=\rho$, $\omega$, or $\phi$, are the
meson-nucleon form factors.

The following parametrization of these form factors is chosen for GKex :
\begin{align}
F_{1}^{\alpha,D}(Q^{2})  &  =\frac{\Lambda_{1,D}^{2}}{\Lambda_{1,D}^{2}%
+\tilde{Q}^{2}}\frac{\Lambda_{2}^{2}}{\Lambda_{2}^{2}+\tilde{Q}^{2}%
},\label{FM_FAFD_GKEX}\\
F_{2}^{\alpha,D}(Q^{2})  &  =\frac{\Lambda_{1,D}^{2}}{\Lambda_{1,D}^{2}%
+\tilde{Q}^{2}}\left(  \frac{\Lambda_{2}^{2}}{\Lambda_{2}^{2}+\tilde{Q}^{2}%
}\right)  ^{2},\nonumber\\[1ex]
F_{1}^{\phi}(Q^{2})  &  =F_{1}^{\alpha}\left(  \frac{Q^{2}}{\Lambda_{1}%
^{2}+Q^{2}}\right)  ^{1.5}\ ,\quad F_{1}^{\phi}(0)=0,\nonumber\\
F_{2}^{\phi}(Q^{2})  &  =F_{2}^{\alpha}\left(  \frac{\Lambda_{1}^{2}}%
{\mu_{\phi}^{2}}\frac{Q^{2}+\mu_{\phi}^{2}}{\Lambda_{1}^{2}+Q^{2}}\right)
^{1.5},\nonumber\\
\text{with}~\tilde{Q}^{2}  &  =Q^{2}\frac{\ln\left[  (\Lambda_{D}^{2}%
+Q^{2})/\Lambda_{\mathrm{QCD}}^{2}\right]  }{\ln(\Lambda_{D}^{2}%
/\Lambda_{\mathrm{QCD}}^{2})}\ .\nonumber
\end{align}

\begin{figure}
[ptb]
\begin{center}
\includegraphics[
trim=0.000000in 0.000000in 0.000788in 0.000000in,
height=5.1915in,
width=5.3835in
]%
{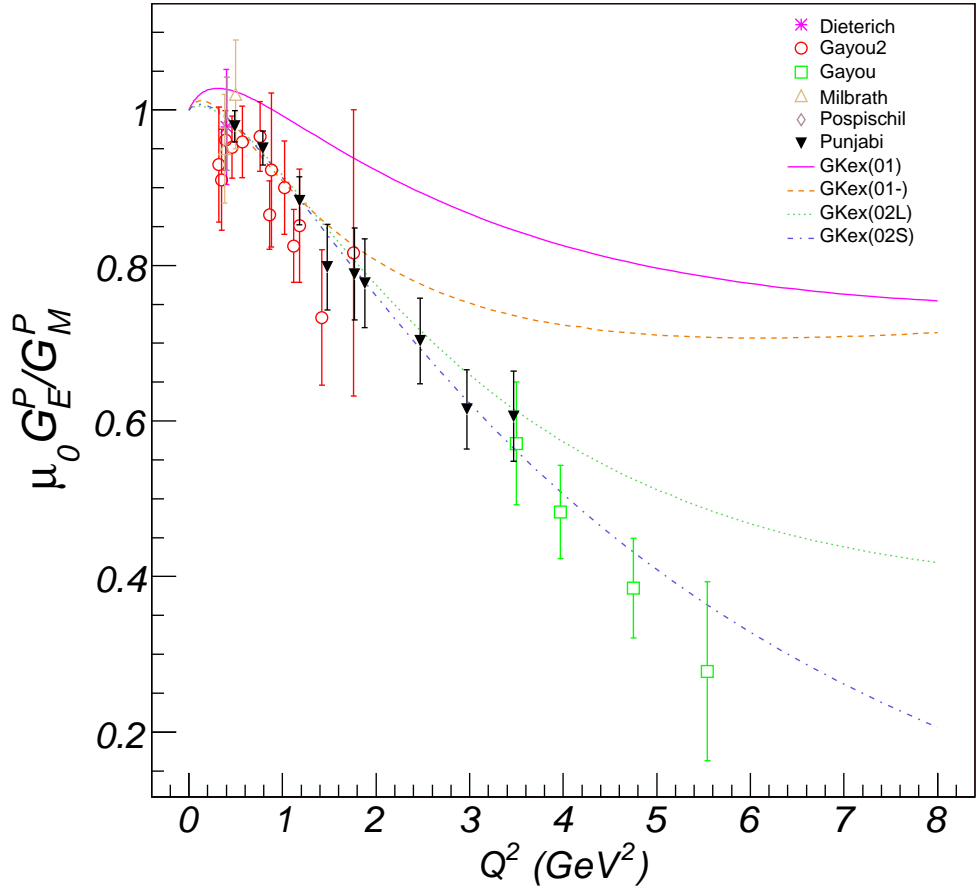}%
\caption{(Color Online)R$_{p}$, the ratio $\mu_{p}G_{p}^{E}/G_{p}^{M}$.
Comparison of fits using the GKex model with the data. The experimental points
used are taken from Refs.: Dieterich \cite{Dieterich:2000mu}, Gayou2
\cite{Gayou:2001qt}, Gayou \cite{Gayou:2001qd}, Milbrath
\cite{Milbrath:1997de}, Pospischil \cite{Pospischil:2001pp} and Punjabi
\cite{Punjabi:2005wq}}%
\label{PL_GKEX_PHYS_COMP}%
\end{center}
\end{figure}

With this formulation there are unknown 8 meson coupling constants, 4 cutoff
masses, one magnetic moment and a single normalization constant, all of which
should be determined from the fits to the experimental data. Fits to the
experimental data points were made using different sets of data, some of which
excluded the controversial high $G_{E}^{p}/G_{M}^{p}$ measured previously by
Rosenbluth separation method. The fits with different data sets were labeled
GKex(01), GKex(01-), GKex(02S) and GKex(02L). The values of the fitted
parameters are listed in the Table I in Ref.\cite{Lomon:2002jx}. Figure
(\ref{PL_GKEX_PHYS_COMP}) shows that the model GKex(02S) describes the
fall-off of $G_{E}^{p}/G_{M}^{p}$ with $Q^{2}$, in contrast with GKex(01) and
GKex(01-), which stay almost flat in the considered range of the $Q^{2}$. In
the present work we use all 4 models in our attempt to describe the lattice
data.%

Using the model to reproduce lattice data requires that we make extrapolations
of some of the parameters that depend on the mass of the hadron constituents.
We start by considering the normalizations of the isovector and isoscalar form
factors that depend on the nucleon magnetic moments:%
\begin{align}
F_{2}^{IV}(0)  &  =\kappa_{\nu}=\left(  \mu_{p}-1-\mu_{n}\right)
;\label{FM_ISIV_NORM}\\
F_{2}^{IS}(0)  &  =\kappa_{s}=\left(  \mu_{p}-1+\mu_{n}\right)  ;\nonumber
\end{align}
The magnetic moments have non-trivial dependence upon the pion mass as a
consequence of chiral symmetry. For example the leading dependence on the
quark mass near the chiral limit is in fact non-analytic (i.e. proportional to
$m_{\pi}\sim m_{q}^{1/2}$). As shown by Leinweber et al. in
Ref.\cite{Leinweber:1998ej}, to extrapolate the nucleon magnetic moments for
the mass range accessible in lattice QCD to the physical mass scale, we use
the Pade%
\'{}
approximant derived in Ref.\cite{Leinweber:1998ej}%
\begin{align}
\mu_{p}\left(  m_{\pi}\right)   &  =\frac{3.31}{1+1.37\cdot m_{\pi}+0.452\cdot
m_{\pi}^{2}},\label{FM_NUC_MM_ETRAP}\\
\mu_{n}\left(  m_{\pi}\right)   &  =\frac{-2.39}{1+1.85\cdot m_{\pi
}+0.271\cdot m_{\pi}^{2}}.\nonumber
\end{align}

The dependence of the masses of the vector mesons upon the pion mass was
studied in the work by Leinweber et al. \cite{RHO_MASS}. We use a linear
extrapolation for the vector meson masses, which was shown in
Ref.\cite{RHO_MASS} to provide quite a good approximation to the full mass
function including the LNA and NLNA behavior:%
\begin{align}
m_{v}\left(  m_{\pi}\right)   &  =c_{0}+c_{1}m_{\pi}^{2};
\label{FM_VM_MASS_EQ}\\
m_{v}\left(  m_{\pi}\right)   &  =m_{v}^{phys}+c_{1}\left(  m_{\pi}%
^{2}-(m_{\pi}^{phys})^{2}\right)  ;\nonumber\\
c_{1}  &  =0.4273%
\operatorname{GeV}%
^{-1};\nonumber
\end{align}

The vector-meson nucleon effective coupling constants may also depend on the
mass of the hadron constituents and to describe that we choose the following
extrapolation forms%
\begin{equation}
g_{i}^{\alpha}\left(  m_{\pi}^{2}\right)  =g_{i0}^{\alpha}+a_{i}^{l_{\alpha}%
}\left(  m_{\pi}^{2}-(m_{\pi}^{phys})^{2}\right)  +b_{i}^{l_{\alpha}}\left(
m_{\pi}^{4}-(m_{\pi}^{phys})^{4}\right)  \label{FM_ISIV_CC_EXTRAP}%
\end{equation}
where $\alpha=\rho^{\prime}$, $\omega$, $\omega^{\prime}$, $\phi$; $l_{\alpha
}=\{IV$ for $\alpha=\rho^{\prime};IS$ for $\alpha=\omega,\omega^{\prime}%
,\phi\}$ ; $i=1,2$; and $g_{10}^{\alpha}=\frac{g_{\alpha}}{f_{\alpha}}$,
$g_{20}^{\alpha}=\kappa_{\omega}\frac{g_{\alpha}}{f_{\alpha}}$ are the
effective coupling constants at the physical $m_{\pi}$. These are taken from
the fits to the physical data of Ref. \cite{Lomon:2002jx}.

We choose a similar ansatz for the extrapolation of the cut-off masses%
\begin{equation}
\Lambda\left(  m_{\pi}^{2}\right)  =\Lambda_{0}+a^{\Lambda}\left(  m_{\pi}%
^{2}-(m_{\pi}^{phys})^{2}\right)  +b^{\Lambda}\left(  m_{\pi}^{4}-(m_{\pi
}^{phys})^{4}\right)  \label{FM_LAMBDA_EXTRAP}%
\end{equation}

where $\Lambda=\Lambda_{1},\Lambda_{2},\Lambda_{D},\Lambda_{QCD}$ and
$\mu_{\phi}$.

\subsection{Fitting Procedure}

Using the extrapolation forms given in Eqs. (\ref{FM_NUC_MM_ETRAP}%
-\ref{FM_LAMBDA_EXTRAP}), we can fit the GKex form factors given by Eq.
(\ref{FM_ISIV_GKEX}) to the lattice data by varying the coefficients $a$,$\ b$
of relations (\ref{FM_ISIV_CC_EXTRAP}) and (\ref{FM_LAMBDA_EXTRAP}). We
performed the fits using the 4 different sets of physical GKex parameters
reported in Ref.\cite{Lomon:2002jx}.

The form factor calculations in Ref.\cite{Gockeler:2003ay} were carried out
using quenched, non-perturbatively $O(a)$-improved Wilson fermions (clover
fermions), for three different values of the lattice spacing,
$a=\{0.47,0.34,0.26\}%
\operatorname{GeV}%
^{-1}$. For each value of $a$ several sets of pion (or equivalently nucleon)
masses were considered. For each mass set Dirac and Pauli form factors for
both the proton and neutron were calculated at several values of $Q^{2}$. The
typical range for the pion mass used varied from $1.2%
\operatorname{GeV}%
$ to $0.6%
\operatorname{GeV}%
$, with the corresponding nucleon mass ranging from approximately $2%
\operatorname{GeV}%
$ to $1.5%
\operatorname{GeV}%
$. The typical range for $Q^{2}$ was $0.6%
\operatorname{GeV}%
^{2}$ to $2.3%
\operatorname{GeV}%
^{2}$. With the smallest lattice spacing being around $0.05%
\operatorname{fm}%
$ ($\beta=6.4$) and pion mass $580%
\operatorname{MeV}%
$, these calculations represent the present the state of the art.

\begin{figure}
[ptb]
\begin{center}
\includegraphics[
height=6.698in,
width=5.3731in
]%
{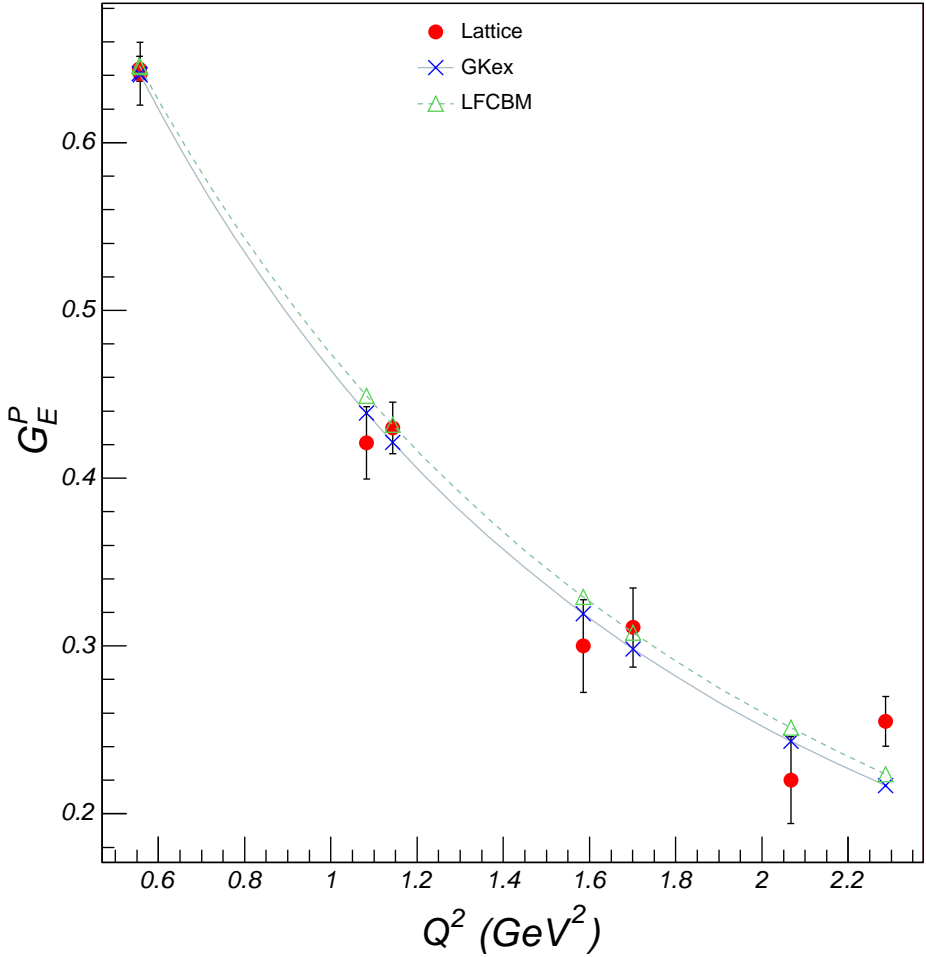}%
\caption{(Color online)GKex(01-) fit[solid] to QCDSF data for $G_{E}%
^{P}(in~units~of~e)$ for a lattice spacing $a=0.26~\operatorname{GeV}^{-1}$,
$M_{N}=2.20~\operatorname{GeV}$ and $m_{\pi}=1.24~\operatorname{GeV}$. LFCBM
fits[dashed] are also shown for comparison.}%
\label{PL_6_4_1_PROT_GE}%
\end{center}
\end{figure}

\begin{figure}
[ptbptb]
\begin{center}
\includegraphics[
height=6.698in,
width=5.3731in
]%
{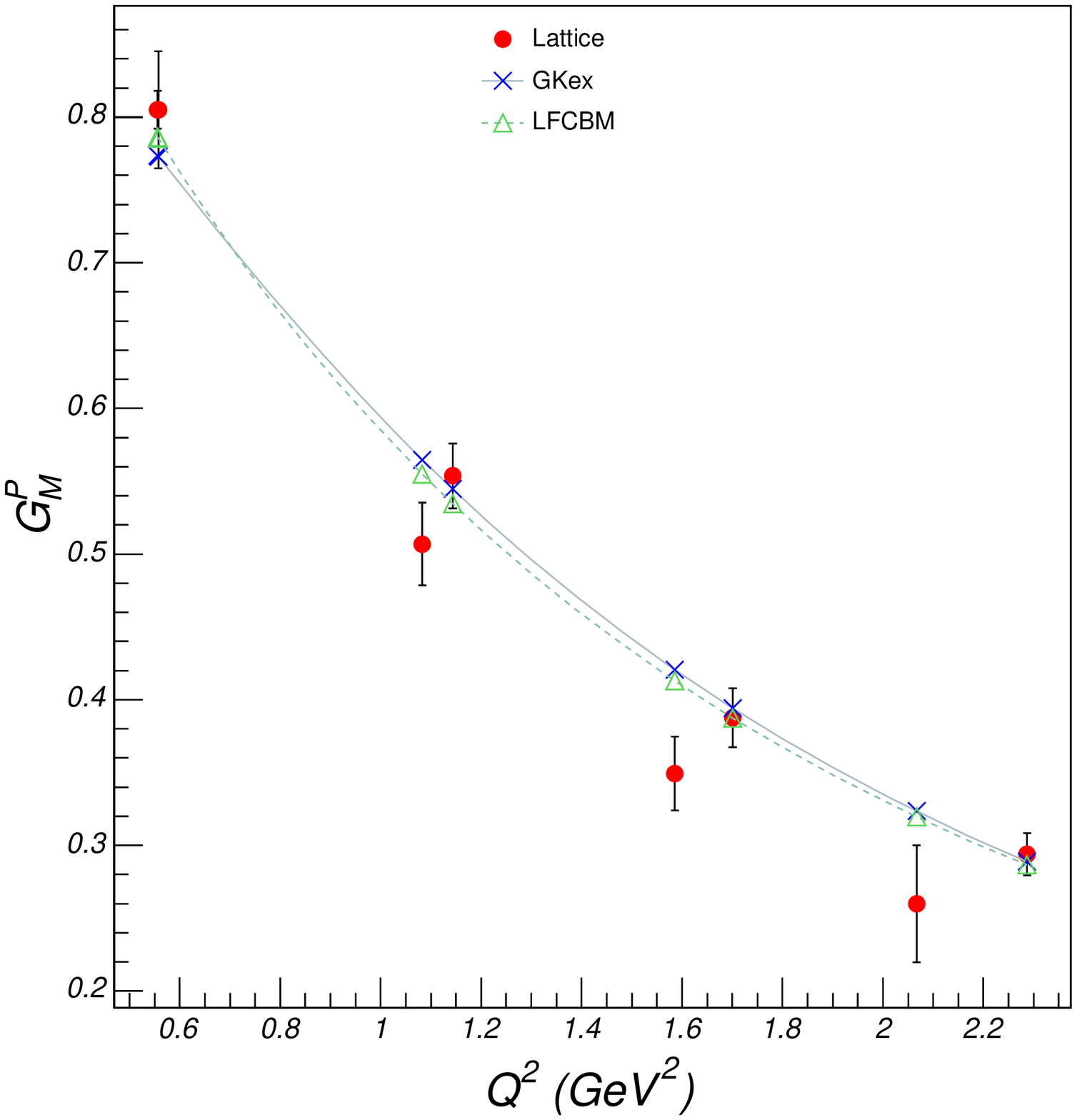}%
\caption{(Color online)GKex(01-) fit[solid] to QCDSF data for $G_{M}%
^{P}(in~units~of~e/(2M_{N}^{Physical}))$ for a lattice spacing
$a=0.26~\operatorname{GeV}^{-1}$, $M_{N}=2.20~\operatorname{GeV}$ and $m_{\pi
}=1.24~\operatorname{GeV}$. LFCBM fits[dashed] are also shown for comparison.}%
\label{PL_6_4_1_PROT_GM}%
\end{center}
\end{figure}

\begin{figure}
[ptbptbptb]
\begin{center}
\includegraphics[
height=6.698in,
width=5.3731in
]%
{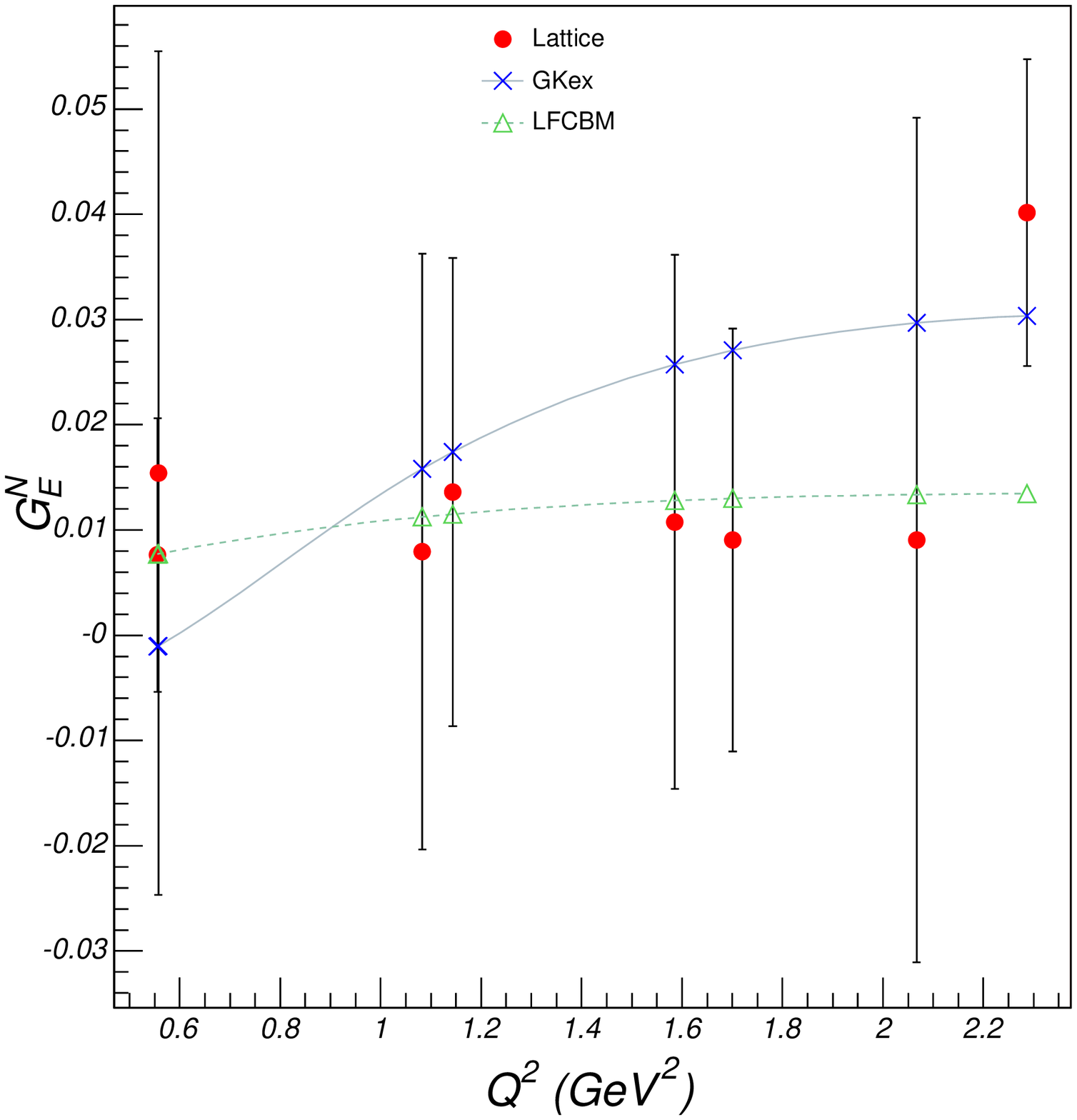}%
\caption{(Color online) GKex(01-) fit[solid] to QCDSF data for $G_{E}%
^{N}(in~units~of~e)$ for a lattice spacing $a=0.26~\operatorname{GeV}^{-1}$,
$M_{N}=2.20~\operatorname{GeV}$ and $m_{\pi}=1.24~\operatorname{GeV}$. LFCBM
fits[dashed] are also shown for comparison.}%
\label{PL_6_4_1_NEUT_GE}%
\end{center}
\end{figure}

\begin{figure}
[ptbptbptbptb]
\begin{center}
\includegraphics[
height=6.698in,
width=5.3731in
]%
{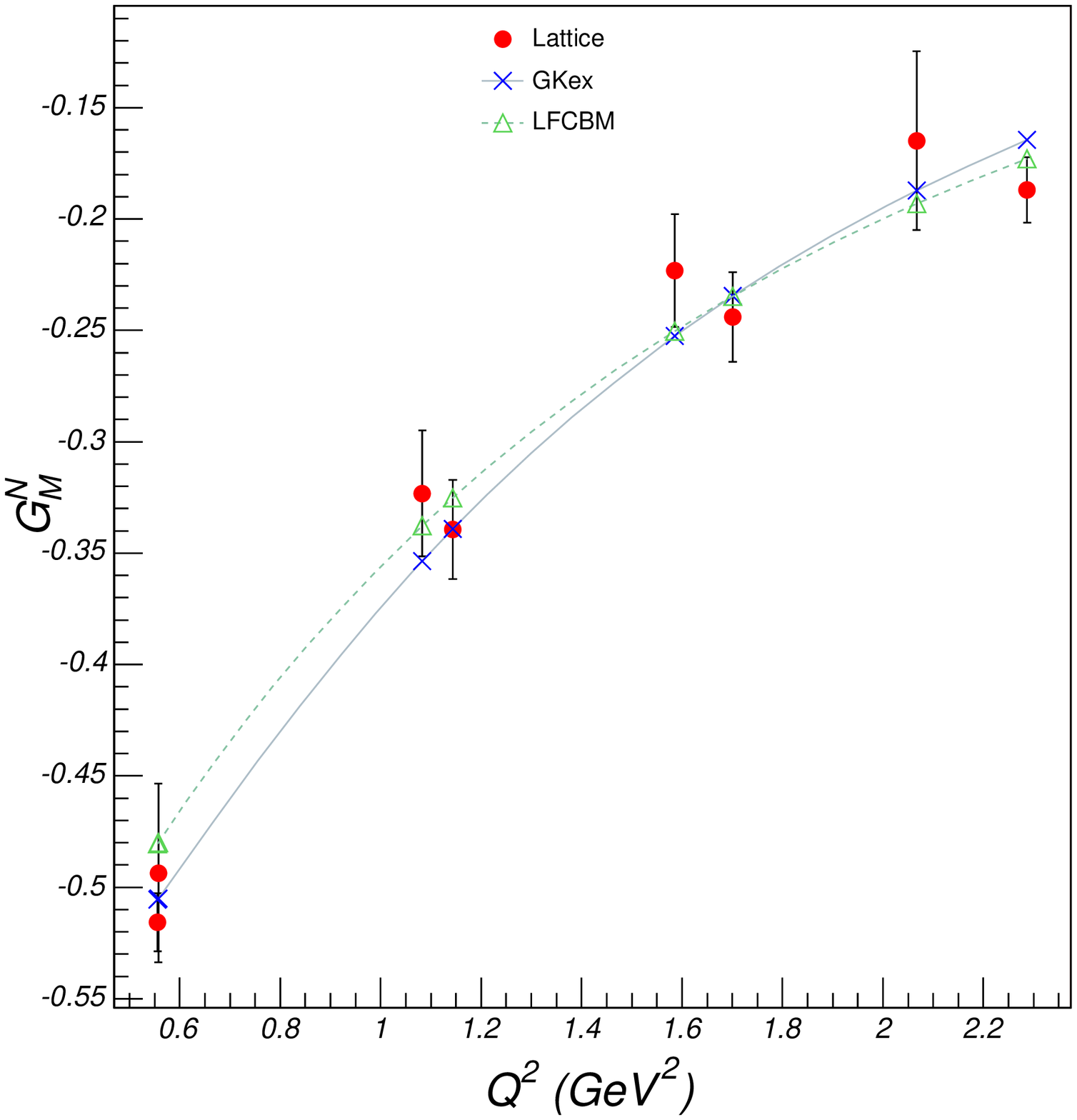}%
\caption{(Color online) GKex(01-) fit[solid] to QCDSF data for $G_{M}%
^{N}(in~units~of~e/(2M_{N}^{Physical}))$ for a lattice spacing
$a=0.26~\operatorname{GeV}^{-1}$, $M_{N}=2.20~\operatorname{GeV}$ and $m_{\pi
}=1.24~\operatorname{GeV}$. LFCBM fits[dashed] are also shown for comparison.}%
\label{PL_6_4_1_NEUT_GM}%
\end{center}
\end{figure}

We fitted the lattice data points for all three lattice spacings available
using the Minuit package of CERN's Root framework \cite{ROOT}. The resulting
fits for the smallest lattice spacing $a=0.26~%
\operatorname{GeV}%
^{-1}$ with $120$ data points are shown in the Figs. \ref{PL_6_4_1_PROT_GE},
\ref{PL_6_4_1_PROT_GM}, \ref{PL_6_4_1_NEUT_GE} and \ref{PL_6_4_1_NEUT_GM},
where the corresponding fits using the LFCBM from our earlier work
\cite{Matevosyan:2005fz} are shown for comparison.%

The resulting $\chi^{2}$ and the fitting parameters for lattice spacing
$a=0.26~%
\operatorname{GeV}%
^{-1}$ are summarized in the Table I, below. For a comparison, the $\chi
^{2}=81$ for the LFCBM fit.%

\begin{gather*}
\text{TABLE I.~GKex fitting parameters and }\chi^{2}\text{ for lattice spacing
}a=0.26~%
\operatorname{GeV}%
^{-1}\text{.}\\%
\begin{tabular}
[c]{ccccc}\hline\hline
& GKex(01) & GKex(01-) & GKex(02L) & GKex(02S)\\\hline\hline
$\chi^{2}$ & 185 & 103 & 671 & 217\\\hline
a$_{1}^{IV}$~ & -1.80(16) & -2.35 & -1.7(2) & -1.86(17)\\
b$_{1}^{IV}$ & 0.46(11) & 0.81 & 0.54(14) & 0.39(13)\\
a$_{2}^{IV}$ & -11.9(5) & -55.2 & -1(1) & -10.9(5)\\
b$_{2}^{IV}$ & 2.98(36) & 18 & -0.34(76) & 2.34(38)\\\hline
a$_{1}^{IS}$ & -0.99(18) & -1.61 & -0.39(9) & -0.58(1)\\
b$_{1}^{IS}$ & -0.21(14) & 0.28 & 0.06(6) & -0.078(73)\\
a$_{2}^{IS}$ & 8.6(18) & 3.1 & 2.53(58) & -0.32(14)\\
b$_{2}^{IS}$ & 2.6(14) & -0.44 & 0.62(41) & 0.1(1)\\\hline
a$^{\Lambda}$ & 0.034(38) & -0.19 & 0.3(1) & 0.035(42)\\
b$^{\Lambda}$ & -0.10(3) & 0.065(33) & -0.14(6) & -0.12(3)\\\hline
\end{tabular}
\end{gather*}

\subsection{Results}

As one can see from Table I, the best fit to the data is obtained using the
GKex(01-) model, even though one is inclined to believe that GKex(02S) gives
the best description of the nucleon structure, since it exhibits the rapid
decrease with $Q^{2}$ of the experimentally measured ratio, $G_{E}^{P}%
/G_{M}^{P}$. One can see this from our Fig.~\ref{PL_GKEX_PHYS_COMP} as well as
the original work of Ref.~\cite{Lomon:2002jx}. We also note that our attempts
to fit the data using only the lowest order polynomial forms in $m_{\pi}$ of
the coupling constants (\ref{FM_ISIV_CC_EXTRAP}), (\ref{FM_LAMBDA_EXTRAP}) did
not yield satisfactory results. Indeed we had to include 10 fitting parameters
for successful extrapolations.

\section{Conclusion}

We have explored the dependence of the nucleon electromagnetic form factors on
quark mass, using recent lattice QCD simulations from the QCDSF group. Since
the VMD approach has been widely used to describe the experimental data at
high $Q^{2}$ (a region of special phenomenological interest at the present
time), we use a modern version of the VMD model, namely the
Gari-Kr\"{u}mpelmann model as implemented by Lomon \cite{Lomon:2002jx}, and
extend it in a natural way. Starting with the existing fit to the experimental
data we find that it is possible to describe the lattice simulations quite
well. However, it was necessary to allow some 10 parameters to vary smoothly
with the pion mass in order to do so. In comparison, the LFCBM produced a fit
of similar quality with only two parameters varied. As a result we are led to
the conclusion that VMD\ is not suitable as a method of chiral extrapolation.

\section{Acknowledgments}

This work was supported in part by the U.S. National Science Foundation (Grant
Number 0140300), the Southeastern Universities Research Association(SURA), DOE
grant DE-AC05-84ER40150, under which SURA operates Jefferson Lab, and also DOE
grant DE-FG02-97ER41014. G.~A.~M. thanks Jefferson Lab for its hospitality
during the course of this work. H. H. M. thanks Jerry Draayer for his support
during the course of the work, the Graduate School of Louisiana State
University for a fellowship partially supporting his research, G. Pogosyan and
S. Vinitsky for their support at the Joint Institute of Nuclear Research.

\end{document}